# TherMod Communication: Low Power or Hot Air?

Christiana Chamon
*Department of Electrical and Computer Engineering*
*Virginia Tech*
*1185 Perry St*
*Blacksburg, VA 24060*
*ccgarcia@vt.edu*

The Kirchhoff-Law-Johnson-Noise (KLJN) secure key exchange scheme leverages statistical physics to enable secure communication with zero average power flow in a wired channel. While the original KLJN scheme requires significant power for operation, a recent wireless modification, TherMod, proposed by Basar claims a "low power" implementation. This paper critically examines this claim. We explain that the additional components inherent in Basar's wireless adaptation substantially increase power consumption, rendering the "low power" assertion inappropriate. Furthermore, we clarify that the security claims of the original KLJN scheme do not directly translate to this wireless adaptation, implying significant security breach. Finally, the scheme looks identical one of the stealth communicators from 2005, which was shown not to be secure.

*Keywords*: unconditional security; wireless modification; power consumption.

## 1. Introduction

The Kirchhoff-Law-Johnson-Noise (KLJN) secure key exchange scheme, introduced by Kish in 2005 [1-3], represents a novel approach to secure communication by exploiting the principles of statistical physics [1-66]. Unlike traditional cryptographic methods that rely on computational complexity, KLJN uses thermal noise generated by resistors to establish a secure key exchange over a wired channel with zero average power flow in the information channel. The core of the KLJN scheme, illustrated in Fig. 1, consists of resistors, switches, and noise generators that create an unconditionally secure system [67] primarily based on the 2nd law of thermodynamics.

Recently, Basar proposed a wireless modification of the KLJN scheme, referred to as the TherMod scheme [4], claiming it achieves "low power" communication. This claim has significant implications for applications where energy efficiency is critical, such as IoT devices and wireless sensor networks. However, the practical realization of the KLJN system introduces complexities that challenge the validity of the "low power" assertion, which the wireless expansion further enhances. This paper critically evaluates Basar's claim, analyzing the power consumption of both the original KLJN scheme and its wireless adaptation. Additionally, we investigate whether the security guarantees of the wired KLJN scheme are upheld in the wireless context, identifying "killer" vulnerabilities, because it has been well known that the KLJN scheme can offer security only in the no-wave limit (quasi static limit) [5-7]. Furthermore, the scheme in [4] is itself essentially unsecure because Alice and Bob do not form parallel resistor pairs to manifest a secure key exchange. The Basar scheme looks like one of the stealth communicators from 2005, which were shown not to be secure [8].

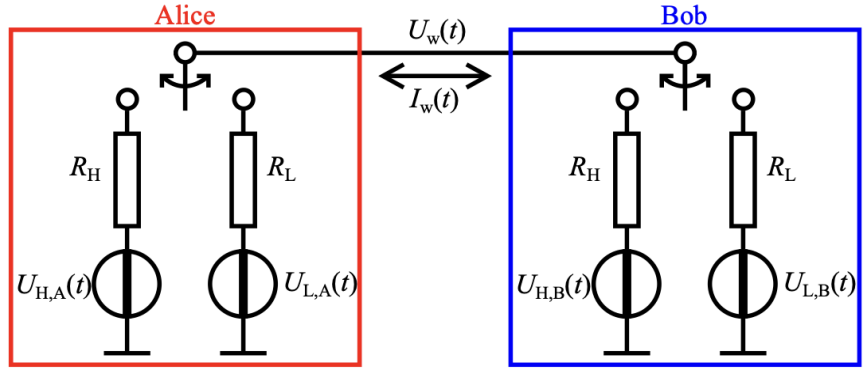

Fig. 1. The core of the KLJN scheme [1-3]. Communicating parties Alice and Bob are connected via a wire. The wire voltage and current are denoted as $U_w(t)$ and $I_w(t)$, respectively. Alice and Bob have identical pairs of resistors $R_H$ and $R_L$ ($R_H > R_L$) that are randomly selected and connected to the wire at the beginning of the bit exchange period. The statistically independent thermal noise voltages $U_{H,A}(t)$, $U_{L,A}(t)$, $U_{H,B}(t)$, and $U_{L,B}(t)$ represent the noise voltages of the resistors $R_H$ and $R_L$ of Alice and Bob, respectively.

## 2. Background and Methodology

### 2.1.   *The KLJN Scheme*

The KLJN scheme operates by leveraging Johnson-Nyquist noise, which arises from the thermal agitation of charge carriers in resistors. In the KLJN setup, the two communicating parties (Alice and Bob) each randomly select one of two resistors ($R_H$ or $R_L$) and connect them to the shared wire. The thermal noise generated by these resistors is measured as voltage and current fluctuations across the channel. By comparing the noise characteristics, Alice and Bob establish a shared secret key. The security of the scheme relies on the indistinguishability of the noise profiles to an eavesdropper (Eve), ensuring unconditional security [1].

The power efficiency of the KLJN scheme is often highlighted due to the zero average power flow in the information channel. However, this does not account for the power consumed by absolutely necessary auxiliary components, such as random number generators, switches, and measurement devices.

### 2.2.   *Basar's Wireless Adaptation (TherMod)*

Basar's TherMod scheme [4] is trying adapt the KLJN principles to a wireless medium, as depicted in Fig. 2. This modification replaces the wired channel with a wireless link, using antennas to transmit and receive noise signals. The TherMod scheme aims to retain the security and power efficiency of the original KLJN system while enabling wireless communication. Basar claims that this adaptation achieves "low power" operation, making it suitable for energy-constrained applications. This mistake is probably due to error of neglecting the power requirements of the devices that run the KLJN system. This power is significant, particularly if one compares it with the very slow data transport of the KLJN scheme that involves amplification and statistical signal processing to extract just a single exchanged bit.

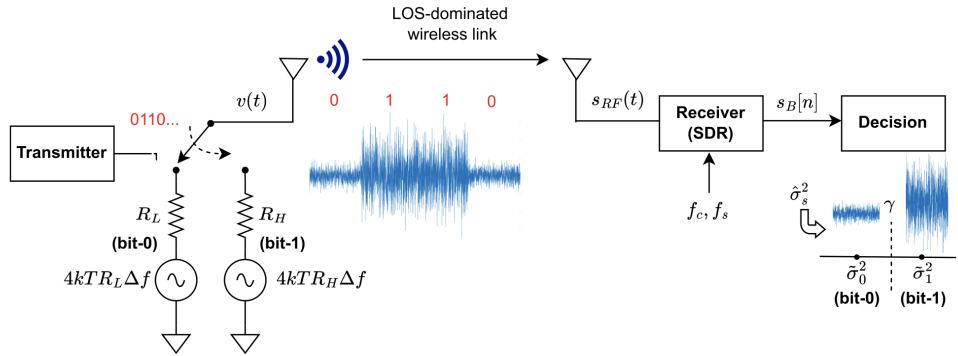

Fig. 2. The core of the proposed TherMod scheme, illustrating the wireless transmission of thermal noise signals using antennas and additional signal processing components [4].

### 2.3. *Methodology*

To evaluate Basar's claim, we analyze the power consumption of both the KLJN and TherMod schemes. We consider the energy requirements of all components, including those not directly involved in the information channel. For the TherMod scheme, we examine the additional components introduced by the wireless medium, such as amplifiers and signal processing units. We also assess the security implications of the wireless adaptation by identifying potential vulnerabilities not present in the wired KLJN scheme.

## 3. Analysis and Refutation of the "Low Power" Claim

### 3.1. *Power Consumption in the KLJN Scheme*

While the KLJN scheme achieves zero average power flow in the information channel, this metric does not apply when assessing the system's overall energy efficiency. The KLJN setup requires several power-consuming components:

- **Random Number Generators**: These produce the noise signals necessary for key exchange, consuming significant energy to ensure high entropy.
- **Switches**: These alternate between resistor values ($R_H$ or $R_L$), requiring power for actuation and control.
- **Measurement Devices**: These amplify voltage and current fluctuations, necessitating sensitive electronics with non-negligible power demands.

- Statistical evaluation of the mean-square voltages and differentiation between secure and non-secure levels.

These components operate continuously during key exchange, resulting in substantial power consumption despite the zero net energy transfer in the channel and thus undermining any claims of inherent energy efficiency.

### 3.2. *Power Consumption in the TherMod Scheme*

Basar's TherMod scheme introduces additional components to enable wireless communication, further increasing power consumption. Key differences include:

- **Signal Amplification**: Wireless transmission requires amplifiers to boost the noise signals to levels suitable for propagation over a wireless medium. These amplifiers consume significant power, especially in environments with high noise or interference.
- **Sophisticated Measurement Techniques**: The wireless medium introduces signal degradation due to path loss, fading, and interference. To accurately measure noise characteristics, the TherMod scheme requires advanced signal processing and error correction, both of which are power-intensive.
- **Antenna Systems**: The use of antennas for transmission and reception adds to the energy budget, particularly for maintaining signal integrity over varying distances.

Our analysis indicates that the power consumption of the TherMod scheme is substantially higher than that of the wired KLJN scheme. For example, a typical RF amplifier used in wireless communication can consume significantly more power than wired measurement circuits. This discrepancy directly contradicts Basar's "low power" claim.

### 3.3. *Quantitative Comparison*

To quantify the power difference, consider a simplified model. In the KLJN scheme, assume the auxiliary components (random number generators, switches, and measurement devices) consume $P_{KLJN}$ during operation, where $P_{KLJN}$ denotes the power consumed by the KLJN scheme. In the TherMod scheme, the addition of an RF amplifier and advanced signal processing increases the total power consumption to $P_{TherMod} = P_{KLJN} + P_{amp} + P_{proc}$, where $P_{TherMod}$ denotes the total power consumed by the TherMod scheme, $P_{amp}$ denotes the power consumed by the RF amplification, and $P_{proc}$ denotes the power consumed by the advanced signal processing. This represents a significant increase, rendering the "low power" label inaccurate.

## 4. Security Considerations

### 4.1. *On "Low Power"*

The security of the original KLJN scheme relies on the physical properties of thermal noise and the wired channel's controlled environment. An eavesdropper attempting to intercept the key must distinguish between noise profiles generated by different resistor combinations, and this is impossible [2].

In the wireless TherMod scheme, the security landscape changes significantly:

- **Environmental Noise**: The wireless medium is susceptible to external noise and interference, which can mask or alter the thermal noise signals, potentially compromising the indistinguishability of noise profiles.
- **Signal Interception**: Wireless signals are inherently broadcast, making it easier for an eavesdropper to capture the transmitted noise without physical access to the channel.
- **Amplification Artifacts**: The amplification process may introduce artifacts that an eavesdropper could exploit to infer information about the resistor states.

These vulnerabilities suggest that the unconditional security of the wired KLJN scheme does not directly translate to the TherMod scheme. Further research is needed to quantify these risks and develop mitigation strategies.

### 4.2.   *KLJN Security Limitations*

The security of the original KLJN scheme relies on the physical properties of thermal noise and the wired channel's controlled, quasi-static environment. In this regime, often referred to as the no-wave limit, an eavesdropper (Eve) attempting to intercept the key must distinguish between noise profiles generated by different resistor combinations. For the 01 and 10 bit configurations, where Alice and Bob select different resistors (RH and RL), the noise variances are identical, rendering it impossible for Eve to determine the specific bit assignments, thus providing unconditional security. However, if Eve knows 50% of the bits, her 75% correct-guessing probability implies substantial information leakage, reducing the effective secrecy capacity of the KLJN scheme. For a 256-bit key, Eve could correctly guess approximately 192 bits, leaving only 64 bits uncertain, significantly weakening security for general communication. Basar's claim that "it would be still difficult to decrypt messages with 50% compromised bits for long enough keys, such as 256-bit keys" [4] is misleading, as the 75% per-bit accuracy facilitates cryptographic attacks unless non-secure bits are discarded through specific protocols, which Basar does not detail.

### 4.3.   *Basar's Non-Novel BEP Claims*

Basar's security claims for KLJN are further undermined by his reliance on a bit-error probability (BEP) framework that lacks novelty when compared to the earlier work of Saez [9]. Saez quantifies the bit error rate (BER) for Alice and Bob in the KLJN scheme, demonstrating that the error probability in key exchange decreases exponentially with the number of samples, achieving a BER of $10^{-6}$ with sufficient sampling. This work focuses on the reliability of bit exchange between the legitimate parties, not on cryptographic security against an eavesdropper. Basar's BEP analysis, which introduces novel detectors (ND-I and ND-II) to reduce errors through joint voltage and current measurements [4], treads similar ground, presenting improvements in error rates as a significant contribution. However, Saez's earlier demonstration of low BER in KLJN key exchange [9] indicates that Basar's framework is not a new advancement but rather a reiteration of established results. Crucially, neither Saez's nor Basar's BEP analyses address the fundamental security flaw of the 75% eavesdropper guessing probability in the non-secure 00 and 11 cases. Basar's assertion of "unconditional security" for the 01/10 cases [4] is technically accurate but overstated, as it fails to mitigate the information leakage in 50% of the exchanges or propose protocols to exclude non-secure bits, rendering the security claim incomplete and vulnerable to the same limitations identified by Saez.

### 4.4.     *Privacy Amplification and Power Costs*

To mitigate the 75% guessing probability, privacy amplification (PA) can be applied, but it introduces significant trade-offs ignored by Basar [10]. For a correct-guessing probability of 75%, four PA iterations ($k = 4$) are required to reduce Eve's guessing probability to approximately 0.5, ensuring negligible information leakage (e.g., $10^{-8}$ bits per exchange) [10]. However, each iteration halves the key length, resulting in a slowdown factor of $2^4 = 16$, as the final key requires 16 times more raw bits [10]. This process also involves discarding non-secure bits (i.e., 00 and 11 cases), necessitating additional key exchange cycles. Both the slowdown and discarded bits increase power dissipation, as more cycles demand sustained operation of power-intensive components like random number generators and measurement devices (see Section 3.1). For example, if the KLJN scheme consumes $P_{KLJN}$ per cycle (Section 3.3), a 16-fold increase in cycles could escalate power consumption to $16P_{KLJN}$, further contradicting Basar's "low power" claim [4]. Basar's failure to acknowledge these PA-induced costs, critical for achieving security in KLJN at a 75% correct-guessing probability, underscores the misleading nature of his security and power efficiency assertions.

### 4.5.     *TherMod's Insecurity*

In the wireless TherMod scheme, the security landscape deteriorates significantly due to its departure from the no-wave limit, a critical requirement for security in the KLJN scheme [7, 8]. The wireless medium introduces wave reflections, interference, and delays beyond the autocorrelation time of the thermal noise, which eliminate the indistinguishability of noise profiles essential for security. These effects allow an eavesdropper to exploit signal variations caused by environmental factors, such as multipath propagation or external noise, to infer resistor states. Moreover, TherMod's unidirectional nature, where a single transmitter modulates noise variance, lacks the parallel resistor configuration of KLJN that enables secure key exchange through ambiguity in the 01/10 cases. Instead, TherMod resembles the wireless stealth communicator proposed by Kish in 2005 [8], which was shown to be insecure due to detectable modulations in the noise spectrum or reflections that an eavesdropper could exploit.

Basar's claim of "stealth communication" for TherMod is particularly misleading, as it lacks the eavesdropper detection mechanisms central to Kish's original stealth communication framework, such as reflection-based detection in thermal equilibrium [8]. In Kish's stealth communicator, the wireless version was vulnerable because modulated noise signals could be intercepted and analyzed without disturbing the channel, a flaw that persists in TherMod. For example, an eavesdropper measuring the transmitted noise variance in TherMod could distinguish between resistor states ($R_L$ or $R_H$) with high probability, especially with large resistance ratios ($\alpha = R_H/R_L$) and sufficient samples, as Basar's own BEP analysis suggests. Unlike KLJN, where the 01/10 ambiguity provides partial security, TherMod offers no such protection, making it inherently insecure. The similarity to Kish's insecure stealth communicator, combined with the violation of the no-wave limit, constitutes a "killer" vulnerability that invalidates Basar's security claims for TherMod.

### 4.6. *Additional Vulnerabilities*

Additional vulnerabilities in TherMod exacerbate these concerns. The wireless medium is susceptible to environmental noise and interference, which can mask or alter the thermal noise signals, potentially compromising the indistinguishability of noise profiles. Wireless signals are inherently broadcast, making it easier for an eavesdropper to capture the transmitted noise without physical access to the channel. Furthermore, the amplification process required for wireless transmission may introduce artifacts that an eavesdropper could exploit to infer information about the resistor states. These vulnerabilities suggest that the unconditional security of the wired KLJN scheme does not translate to the TherMod scheme, and Basar's failure to quantify these risks or propose mitigation strategies further undermines the credibility of his security assertions.

In summary, Basar's security claims for both KLJN and TherMod are fundamentally flawed. For KLJN, the 75% eavesdropper guessing probability highlights significant information leakage, unaddressed by Basar's BEP-focused detectors (which merely echo Saez's earlier BER results without advancing security) or vague claims about decryption difficulty. For TherMod, the violation of the no-wave limit and its resemblance to Kish's insecure stealth communicator eliminate any security guarantees, rendering the "stealth" claim inappropriate. These findings reinforce the need for rigorous security analyses in KLJN-based schemes and highlight the challenges of extending physical-layer security to wireless environments.

## 5. Conclusion

The Kirchhoff-Law-Johnson-Noise (KLJN) secure key exchange scheme offers an innovative approach to secure communication by leveraging thermal noise. However, claims of "low power" operation, particularly in Basar's wireless TherMod adaptation, are misleading. Our analysis demonstrates that the TherMod scheme's additional components, such as amplifiers and advanced signal processing units, significantly increase power consumption compared to the wired KLJN scheme. Furthermore, the security guarantees of the original KLJN system are not assured in the wireless context due to environmental noise, signal interception risks, and amplification artifacts.

These findings highlight the challenges of achieving low-power, secure communication in wireless environments using KLJN-based schemes. Researchers and practitioners should approach such claims with skepticism and prioritize comprehensive power and security analyses in future proposals. While the KLJN scheme remains a promising concept, its practical implementation, requires careful consideration of energy and security trade-offs. Finally, security in the KLJN scheme exists only in the no-wave limit. Any waves reflections, interference, delays beyond the autocorrelation time eliminate the security. The scheme in [4] is rather imitating the "stealth" thermal noise communicator of Kish [8] which has wireless version, too, and was shown not to be secure.

**References**

[1] C. Chamon and L.B. Kish, Perspective–On the Thermodynamics of Perfect Unconditional Security, *Appl. Phys. Lett.* **119** (2021) 010501.
[2] L.B. Kish, Totally secure classical communication utilizing Johnson (-like) noise and Kirchhoff's law, *Phys. Lett. A* (2006).
[3] L.B. Kish, Enhanced secure key exchange system based on the Johnson(-like) noise, *Fluct. Noise Lett.* (2007).
[4] E. Basar, Communication by Means of Thermal Noise: Toward Networks With Extremely Low Power Consumption, *IEEE Comm.* (2022).
[5] L.B. Kish, D. Abbott, C.G. Granqvist, Critical analysis of the Bennett-Riedel attack on secure cryptographic key distributions via the Kirchhoff-law-Johnson-noise scheme, *PLoS ONE* 8 (2013) e81810; https://doi.org/10.1371/journal.pone.0081810
[6] L.B. Kish, T. Horvath, Notes on Recent Approaches Concerning the Kirchhoff-Law-Johnson-Noise-based Secure Key Exchange, *Phys. Lett. A* 373 (2009) 2858-2868.
[7] L.B. Kish, S.P. Chen, C.G. Granqvist, J. Smulko, Waves in a short cable at low frequencies, or just hand- waving? What does physics say?, invited talk at the 23rd International Conference on Noise and Fluctuations (ICNF 2015), Xian, China, June 2-5, 2015. DOI: 10.1109/ICNF.2015.7288604
[8] L.B. Kish, Stealth communication: Zero-power classical communication, zero-quantum quantum communication and environmental-noise communication, *Appl. Phys. Lett.* 87 (2005), Art. No. 234109.
[9] Y. Saez, L.B. Kish, R. Mingesz, Z. Gingl, and C. Granqvist. Current and voltage based bit errors and their combined mitigation for the Kirchhoff-law-Johnson-noise secure key exchange, *Journ. Comp. Elec.* 13 (2013) 10.1007/s10825-013-0515-2.
[10] T. Horvath, L.B. Kish, and J. Schuer. Effective privacy amplification for secure classical communications. *Elec. Phys. Lett*. 94 (2011) 28002.
[11] C. Chamon, S. Ferdous, and L.B. Kish, Random number generator attack against the Kirchhoff-law-Johnson-noise secure key exchange protocol, *arXiv* preprint, https://arxiv.org/abs/2005.10429, 2020.
[12] C. Chamon, S. Ferdous, and L. B. Kish, Deterministic random number generator attack against the Kirchhoff-law-Johnson-noise secure key exchange protocol, *Fluct. Noise Lett.* 20(5) (2021)
[13] C. Chamon, S. Ferdous, and L. B. Kish, "Statistical random number generator attack against the Kirchhoff-law-Johnson-noise secure key exchange protocol, *Fluct. Noise Lett.* (2021).
[14] C. Chamon, S. Ferdous, and L. B. Kish, Nonlinearity attack against the Kirchhoff-law- Johnson-noise secure key exchange protocol, *Fluct. Noise Lett.* (2021).
[15] L. B. Kish, *The Kish Cypher: The Story of KLJN for Unconditional Security*, New Jersey: World Scientific (2017).
[16] A. Cho, *Simple* noise may stymie spies without quantum weirdness, *Science* 309(5744) (2005) 2148.
[17] L.B. Kish and C.G. Granqvist, On the security of the Kirchhoff-law-Johnson-noise (KLJN) communicator, *Quant. Inf. Proc.* 13(10) (2014) 2213-2219.
[18] L. B. Kish and T. Horvath, Notes on recent approaches concerning the Kirchhoff-law–Johnson-noise based secure key exchange, *Phys. Lett. A* 373 (2009) 2858-2868.
[19] G. Vadai, R. Mingesz, and Z. Gingl, Generalized Kirchhoff-law-Johnson-noise (KLJN) secure key exchange system using arbitrary resistors, *Sci. Rep.* 5 (2015).
[20] S. Ferdous, C. Chamon, and L. B. Kish, Comments on the "generalized" KJLN key exchanger with arbitrary resistors: power, impedance, security, *Fluct. Noise Lett.*, 20 (2020) 2130002.

[21] L. B. Kish and C. G. Granqvist, Random-resistor-random-temperature Kirchhoff-law-Johnson-noise(RRRT -KLJN) key exchange, *Met. Meas. Sys.* 23 (2016) 3-11.
[22] J. Smulko, Performance analysis of the 'intelligent' Kirchhoff's-law–Johnson-noise secure key exchange, *Fluct. Noise Lett.* 13 (2014) 1450024.
[23] R. Mingesz, Z. Gingl, and L. B. Kish, Johnson(-like)-noise-Kirchhoff-loop based secure classical communicator characteristics, for ranges of two to two thousand kilometers, via model-line, *Phys. Lett. A* 372 (2008) 978–984.
[24] R. Mingesz, L. B. Kish, Z. Gingl, C. G. Granqvist, H. Wen, F. Peper, T. Eubanks, and G. Schmera, Unconditional security by the laws of classical physics, *Met. Meas. Sys.* 20 (2013) 3–16.
[25] Y. Saez and L. B. Kish, Errors and their mitigation at the Kirchhoff-law-Johnson-noise secure key exchange, *PLoS ONE* 8 (2013) 11.
[26] R. Mingesz, G. Vadai, and Z. Gingl, What kind of noise guarantees security for the Kirchhoff-loop-Johnson-noise key exchange? *Fluct. Noise Lett.* 13(3) (2014) 1450021.
[27] Y. Saez, L. B. Kish, R. Mingesz, Z. Gingl, and C. G. Granqvist, Bit errors in the Kirchhoff-law-Johnson-noise secure key exchange, *Intern. Journ. Mod. Phys: Conf. Ser* 33 (2014) 1460367.
[28] Z. Gingl and R. Mingesz, Noise properties in the ideal Kirchhoff-law-Johnson-noise secure communication system, *PLoS ONE* 9(4) (2014).
[29] P.L. Liu, A key agreement protocol using band-limited random signals and feedback, *IEEE Journ. Light. Tech.* 27(23) (2009) 5230-5234.
[30] L. B. Kish and R. Mingesz, Totally secure classical networks with multipoint teleclonig (teleportation) of classical bits through loops with Johnson-like noise, *Fluct. Noise Lett.* 6(2) (2006) C9–C21.
[31] L.B. Kish, Methods of using existing wirelines (powerlines, phonelines, internetlines) for totally secure classical communication utilizing Kirchoff's law and Johnson-like noise, *arXiv* preprint, https://arxiv.org/abs/physics/0610201, 2006.
[32] L.B. Kish and F.Peper, Information networks secured by the laws of physics, *IEICE Trans. Fund. Comm. Elec. Inf. Sys.* E95–B(5) (2012) 1501-1507.
[33] E. Gonzalez, L. B. Kish, R. S. Balog, and P. Enjeti, "Information theoretically secure, enhanced Johnson noise based key distribution over the smart grid with switched filters, *PloS One* 8(7) (2013).
[34] E. Gonzalez, L. B. Kish, and R. Balog, Encryption Key Distribution System and Method, U.S. Patent #US9270448B2 (2016) https://patents.google.com/patent/US9270448B2.
[35] E. Gonzalez, R. Balog, R. Mingesz, and L. B. Kish, Unconditional security for the smart power grids and star networks, 23rd International Conference on Noise and Fluctuations (ICNF 2015), Xian, China, June 2-5, 2015.
[36] E.Gonzalez, R.S. Balog, and L.B.Kish, Resourcer equirements and speed versus geometry of unconditionally secure physical key exchanges, *Entropy* 17(4) (2015) 2010–2014.
[37] E. Gonzalez and L. B. Kish, Key exchange trust evaluation in peer-to-peer sensor networks with unconditionally secure key exchange, *Fluct. Noise Lett.* 15 (2016) 1650008.
[38] L. B. Kish and O. Saidi, Unconditionally secure computers, algorithms and hardware, such as memories, processors, keyboards, flash and hard drives, *Fluct. Noise Lett.* 8 (2008) L95–L98.
[39] L. B. Kish, K. Entesari, C. G. Granqvist, and C. Kwan, "Unconditionally secure credit/debit card chip scheme and physical unclonable function," *Fluct. Noise Lett.* 16 (2017) 1750002.
[40] L. B. Kish and C. Kwan, Physical unclonable function hardware keys utilizing Kirchhoff-law-Johnson-noise secure key exchange and noise-based logic, *Fluct. Noise Lett.* 12 (2013) 1350018.
[41] Y. Saez, X. Cao, L. B. Kish, and G. Pesti, Securing vehicle communication systems by the KLJN key exchange protocol, *Fluct. Noise Lett.* 13 (2014) 1450020.
[42] X. Cao, Y. Saez, G. Pesti, and L. B. Kish, On KLJN-based secure key distribution in vehicular communication networks, *Fluct. Noise Lett.* 14 (2015) 1550008.
[43] L. B. Kish and C. G. Granqvist, Enhanced usage of keys obtained by physical, unconditionally secure distributions, *Fluct. Noise Lett.* 14 (2015) 1550007.
[44] P. L. Liu, A complete circuit model for the key distribution system using resistors and noise sources, *Fluct. Noise Lett.* 19 (2020) 2050012.
[45] M. Y. Melhem and L. B. Kish, Generalized DC loop current attack against the KLJN secure key exchange scheme, *Met. Meas. Sys.* 26 (2019) 607-616.
[46] M. Y. Melhem and L. B. Kish, A static-loop-current attack against the Kirchhoff-law-Johnson-noise (KLJN) secure key exchange system, *Appl. Sci.* 9 (2019) 666, 2019.

[47] M.Y. Melhem and L.B. Kish, The problem of information leak due to parasitic loop currents and voltages in the KLJN secure key exchange scheme, *Met. Meas. Sys.* 26 (2019) 37-40.
[48] M. Y. Melhem and L. B. Kish, Man in the middle and current injection attacks against the KLJN key exchanger compromised by DC sources, *Fluct. Noise Lett.* 20(2) (2021) 2150011.
[49] M. Y. Melhem, C. Chamon, S. Ferdous, L. B. Kish, Alternating (AC) Loop Current Attacks against the KLJN Secure Key Exchange Scheme, *Fluct. Noise Lett*. (2021).
[50] P. L. Liu, Re-examination of the cable capacitance in the key distribution system using resistors and noise sources, *Fluct. Noise Lett.* 16 (2017) 1750025.
[51] H. P. Chen, M. Mohammad, and L. B. Kish, Current injection attack against the KLJN secure key exchange, *Met. Meas. Sys.* 23 (2016) 173-181.
[52] G. Vadai, Z. Gingl, and R. Mingesz, Generalized attack protection in the Kirchhoff-law-Johnson-noise key exchanger, *IEEE Access* 4 (2016) 1141-1147.
[53] H.P. Chen, E.Gonzalez, Y. Saez, and L.B. Kish, "Cable capacitance attack against the KLJN secure key exchange," *Information* 6 (2015) 719-732.
[54] L.B. Kish and C.G. Granqvist, "Elimination of a second-law-attack, and all cable-resistance-based attacks, in the Kirchhoff-law-Johnson-noise (KLJN) secure key exchange system, *Entropy* 16 (2014) 5223-5231.
[55] L. B. Kish and J. Scheuer, Noise in the wire: the real impact of wire resistance for the Johnson (-like) noise based secure communicator, *Phys. Lett. A* 374 (2010) 2140-2142.
[56] F. Hao, Kish's key exchange scheme is insecure, *IEE Proc. – Inf. Sec.* 153(4) (2006) 141-142.
[57] L. B. Kish, Response to Feng Hao's paper "Kish's key exchange scheme is insecure," *Fluct. Noise Lett.* 6 (2006) C37-C41.
[58] L. B. Kish, Protection against the man-in-the-middle-attack for the Kirchhoff-loop-Johnson (-like)-noise cipher and expansion by voltage-based security, *Fluct. Noise Lett.* 6 (2006) L57-L63.
[59] L. J. Gunn, A. Allison, and D. Abbott, A new transient attack on the Kish key distribution system, *IEEE Access* 3 (2015) 1640-1648.
[60] L. B. Kish and C. G. Granqvist, Comments on "a new transient attack on the Kish key distribution system", *Met. Meas. Sys.* 23 (2015) 321-331.
[61] L. J. Gunn, A. Allison, and D. Abbott, A directional wave measurement attack against the Kish key distribution system, *Sci. Rep.* 4 (2014) 6461.
[62] H. P. Chen, L. B. Kish, and C. G. Granqvist, On the "cracking" scheme in the paper "a directional coupler attack against the Kish key distribution system" by Gunn, Allison and Abbott, *Met. Meas. Sys.* 21 (2014) 389-400.
[63] H. P. Chen, L. B. Kish, C. G. Granqvist, and G. Schmera, Do electromagnetic waves exist in a short cable at low frequencies? What does physics say? *Fluct. Noise Lett.* 13 (2014) 1450016.
[64] L. B. Kish, Z. Gingl, R. Mingesz, G. Vadai, J. Smulko, and C. G. Granqvist, Analysis of an attenuator artifact in an experimental attack by Gunn–Allison–Abbott against the Kirchhoff-law–Johnson-noise (KLJN) secure key exchange system, *Fluct. Noise Lett.* 14 (2015) 1550011.
[65] S. Ferdous, C. Chamon, and L.B. Kish, Current injection and voltage insertion attacks against the VMG-KLJN secure key exchanger, *Fluct. Noise Lett.* 22 (2023) 2350009.
[66] K. Mohanasundar, S.A. Flanery, C. Chamon, and S.D. Kotikela. Kirchhoff-Law Johnson Noise Meets Web 3.0: A Statistical Physical Method of Random Key Generation for Decentralized Identity Protocols. *arXiv* preprint (2023) arXiv:2312.17113
[67] Y. Liang, H. V. Poor, S. Shamai, "Information theoretic security," *Foundations and Trends in Communications and Information Theory*, vol. 5, pp. 355-380, 2008.

.